\documentstyle[preprint,aps]{revtex}
\newcommand{\be}{\begin{equation}}
\newcommand{\ee}{\end{equation}}
\newcommand{\bea}{\begin{eqnarray}}
\newcommand{\eea}{\end{eqnarray}}
\newcommand{\bm}{\boldmath}

\newcommand{\mb}{\mbox}
\def\scri{\hbox{${\cal J}$\kern -.645em 
{\raise.57ex\hbox{$\scriptscriptstyle(\ $}}}}

\newcommand{\jmp}{{\em J. Math. Phys.\/}}

\tighten

\begin{document}
\draft
\title{On the Equivalence of Brans-Dicke and Einstein Theories with 
Static Black Holes}
\author{Eloy Ay\'on}
\address{Grupo de F\'\i sica Te\'orica, Centro de Matem\'atica y F\'\i sica
Te\'orica, ICIMAF\\
Calle E $\#$ 309, esq. a 15, CP 10400, Ciudad Habana, Cuba}
\maketitle

\begin{abstract}
A ``no-hair'' theorem for the Brans-Dicke field, in the domain of outer
communications of a static black hole, is established, when the trace of the
energy-momentum tensor, $T$, is a definite-sign scalar. This implies that
Brans-Dicke theory of gravitation, for matter with $T$ of definite sign, is
equivalent to Einstein gravity with the constraint $T=0$.
\end{abstract}




~\newline

The Brans-Dicke theory of gravitation is an attempt to improve General
Relativity from the standpoint of Mach principle \cite{BransDicke,Brans62}.
It is not a completely geometrical theory since gravitational effects are
described not only by the metric but also by a scalar field $\Psi $, which
plays a role analogous to the reciprocal of the local gravitational
constant, $\Psi \sim 1/G$. However, in the presence of stationary black
holes, it was conjectured \cite{Penrose70} the equivalence of this theory to
Einstein gravity. This fact was established in \cite{Hawking72}, and
independently in \cite{BekensPhD,Bekens72c} for static or circular black
holes with matter having vanishing trace of the energy-momentum tensor, 
$T\equiv T_\alpha ^{~\alpha }=0$, {\em e.g.\/}, for electromagnetic fields,
neutrino fields, conformal scalar fields. More recently, for spherical symmetric 
static black holes, the equivalence has been showed for scalar models having 
positive energy \cite{Beke95}, and has been rederived for the vacuum case 
\cite{Saa96a}.

The aim of this letter is to extend the previous results to more general
matter systems in static black holes. In essence, we concentrate our
attention in matter for which $T$ has definite sign, {\em i.e.\/}, it is
non-positive, or non-negative; since the scalar models studied in 
\cite{Beke95} belong to this class ($T\leq 0$).

The field equations for the Brans-Dicke theory are 
\begin{equation}
\Psi G_{\mu \nu }=\frac{8\pi }{c^4}T_{\mu \nu }+\frac \omega \Psi (\nabla
_\mu \Psi \nabla _\nu \Psi -\frac 12g_{\mu \nu }\nabla _\alpha \Psi \nabla
^\alpha \Psi )+\nabla _\mu \nabla _\nu \Psi -g_{\mu \nu }\Box \Psi ,
\label{eq:B-D}
\end{equation}
\begin{equation}
\Box \Psi =\frac 1{2\omega +3}\frac{8\pi }{c^4}T,  \label{eq:B-Df}
\end{equation}
plus matter equations, where $\omega \geq 6$ is a coupling constant 
\cite{BransDicke}. From (\ref{eq:B-D}) and (\ref{eq:B-Df}) it follows that 
the Brans-Dicke-matter system with $\Psi =\text{const.}$ is equivalent to 
the Einstein-matter system with the constraint $T=0$. We shall establish 
a ``no-hair'' theorem for the Brans-Dicke field $\Psi $, in the domain of
outer communications $\langle \!\langle 
\hbox{${\cal J}$\kern -.645em
{\raise.57ex\hbox{$\scriptscriptstyle (\ $}}}\rangle \!\rangle $ of a static
black hole, for which $T$ is a definite-sign scalar, {\em i.e.\/}, if 
$T\leq0$, or $T\geq 0$, then $\Psi =\Psi _o=\text{const.}$, where 
$\Psi _o$ is the asymptotic value of $\Psi $.

In a static black hole, the Killing field {\boldmath$k$} coincides with the
null generator of the event horizon ${{\cal {H}}}^{+}$ and is time-like and
hypersurface orthogonal in all of $\langle \!\langle 
\hbox{${\cal J}$\kern -.645em
{\raise.57ex\hbox{$\scriptscriptstyle (\ $}}}\rangle \!\rangle $. This,
together with simply connectedness of $\langle \!\langle 
\hbox{${\cal J}$\kern -.645em
{\raise.57ex\hbox{$\scriptscriptstyle (\ $}}}\rangle \!\rangle $ 
\cite{ChrWald95}, implies the existence of a global coordinates system, 
$(t,x^i)$ $i=1,2,3$, in $\langle \!\langle 
\hbox{${\cal J}$\kern -.645em
{\raise.57ex\hbox{$\scriptscriptstyle (\ $}}}\rangle \!\rangle $ 
\cite{Carter87}, where 
$\mbox{\boldmath$k$}=\mbox{\bm$\partial/\partial t$}$ such
that the metric can be expressed by 
\begin{equation}
\mbox{\bm $g$}=-V\mbox{\bm $dt$}^2+\gamma _{ij}\mbox{\bm
$dx$}^i\mbox{\bm $dx$}^j,  \label{eq:static}
\end{equation}
where $V$ and {\boldmath$\gamma $} are $t$-independent, 
{\boldmath$\gamma $}
being positive definite in all of $\langle \!\langle 
\hbox{${\cal J}$\kern -.645em
{\raise.57ex\hbox{$\scriptscriptstyle (\ $}}}\rangle \!\rangle $, and $V$
being a positive function tending to zero in ${{\cal {H}}}^{+}$.

Let us begin first with the case $T\leq 0$. The idea is to assume 
$\Psi|_{\langle \!\langle 
\hbox{${\cal J}$\kern -.645em
{\raise.57ex\hbox{$\scriptscriptstyle (\ $}}}\rangle \!\rangle }\not\equiv
\Psi _o$, and then to obtain a contradiction. Since {\boldmath$k$} is
time-like in all of $\langle \!\langle 
\hbox{${\cal J}$\kern -.645em
{\raise.57ex\hbox{$\scriptscriptstyle (\ $}}}\rangle \!\rangle $, the
elliptical nature of the Brans-Dicke-matter system guaranties that $\Psi $, 
{\boldmath$g$}, and the matter fields are all analytical in appropriate
coordinates \cite{Muller70}. If $\Psi |_{\langle \!\langle 
\hbox{${\cal J}$\kern -.645em
{\raise.57ex\hbox{$\scriptscriptstyle (\ $}}}\rangle \!\rangle }\not \equiv
\Psi _o$, using its analyticity, it can be showed \cite{MullerHRS73,Ayon3}
that if $\Psi ^{-1}(\Psi _o)\neq \emptyset $, then it is a union of
countably many 0-dimensional (point-like), 1-dimensional, 2-dimensional, 
and 3-dimensional analytic sub-manifolds of 
$\langle \!\langle \hbox{${\cal J}$\kern -.645em
{\raise.57ex\hbox{$\scriptscriptstyle (\ $}}}\rangle \!\rangle $. 
Let ${\cal V}\subset \langle \!\langle 
\hbox{${\cal J}$\kern -.645em
{\raise.57ex\hbox{$\scriptscriptstyle (\ $}}}\rangle \!\rangle $ be the open
region bounded by the space-like hypersurface $\Sigma _0$, the space-like
hypersurface $\Sigma _1$, corresponding to the values $0$ and $1$ of $t$
along integral curves of {\boldmath$k$}, and portions of the event horizon 
{\em ${\cal H}^{+}$}, and the spatial infinity $i^o$ respectively.
Analyticity of $\Psi $ together with the connectedness of 
$\langle \!\langle 
\hbox{${\cal J}$\kern -.645em
{\raise.57ex\hbox{$\scriptscriptstyle (\ $}}}\rangle \!\rangle $ implies
that $\Psi |_{{\cal V}}\not \equiv \Psi _o$ if $\Psi |_{\langle \!\langle 
\hbox{${\cal J}$\kern -.645em
{\raise.57ex\hbox{$\scriptscriptstyle (\ $}}}\rangle \!\rangle }
\not \equiv\Psi _o$. Let $f\in C^\infty (I\!\!R\setminus \Psi _o)$ 
be a real function defined by 
\begin{equation}
f(t)=\left\{ 
\begin{array}{cr}
0, & t\leq \Psi _o, \\ 
\exp (-1/(t-\Psi _o)^2)-1, & t>\Psi _o.
\end{array}
\right.  \label{eq:f}
\end{equation}
Such function satisfies 
\begin{equation}
f(\Psi _o)=0,\qquad -1\leq f(t)\leq 0,\qquad \left. f^{\prime }(t)\right|
_{I\!\!R\setminus \Psi _o}\geq 0.  \label{eq:prop}
\end{equation}
Multiplying the Brans-Dicke field equation (\ref{eq:B-Df}) by 
$f\circ \Psi $ and integrating over ${\cal {V}}\setminus\Psi ^{-1}(\Psi _o)$, 
after an integration by parts making use of the fact that $\Psi ^{-1}(\Psi _o)$ 
is a measure zero set, and the Gauss theorem, one arrives at 
\begin{equation}
\int_{\partial {\cal {V}}}f(\Psi )\nabla ^\mu \Psi \,n_\mu d\sigma =\int_{%
{\cal {V}}\setminus \Psi ^{-1}(\Psi _o)}(f^{\prime }(\Psi )\nabla _\mu \Psi
\nabla ^\mu \Psi +\frac{f(\Psi )}{2\omega +3}\frac{8\pi }{c^4}T)dv.
\label{eq:ints}
\end{equation}
The boundary integral over $\Sigma _1$ cancels out that over $\Sigma _0$ at
the left hand side of (\ref{eq:ints}), since $\Sigma _1$ and $\Sigma _0$ are
isometric surfaces. The boundary integral over $i^o\cap\overline{{\cal{V}}}$ 
is zero, because $f(\Psi )$ is zero there (\ref{eq:prop}). The boundary
integral over ${\cal {H}}^{+}\cap \overline{{\cal {V}}}$ vanishes too; 
since {\boldmath$k$} coincides with the normal of ${\cal {H}}^{+}$, 
$\nabla ^\mu\Psi \,n_\mu =\mbox{\bm$\pounds_{\mb{\bm$k$}}$}\Psi =0$ 
by staticity of $\Psi $, and by the bounded
behavior of $f(\Psi )$ (\ref{eq:prop}) the integrand vanishes in this
region. Since there are no contributions at the left hand side of 
(\ref{eq:ints}) the volume integral vanishes, thus 
\begin{equation}
\int_{{\cal {V}}\setminus \Psi ^{-1}(\Psi _o)}(f^{\prime }(\Psi )\gamma
_{ij}\nabla ^i\Psi \nabla ^j\Psi +\frac{f(\Psi )}{2\omega +3}\frac{8\pi }{c^4%
}T)dv=0,  \label{eq:zeros1}
\end{equation}
where the coordinates chosen are the ones of (\ref{eq:static}). From the
properties of $f$, $T$, and {\boldmath$\gamma $} it follows that each term
in the integrand above is non-negative, so that (\ref{eq:zeros1}) is
fulfilled only if these terms vanish identically in 
${\cal {V}}\setminus\Psi ^{-1}(\Psi _o)$, in particular, 
\begin{equation}
\left. f^{\prime }(\Psi )\gamma _{ij}\nabla ^i\Psi \nabla ^j\Psi \right| _{%
{\cal {V}}\setminus \Psi ^{-1}(\Psi _o)}=0.  \label{eq:van}
\end{equation}
The analytical function $E\equiv \gamma _{ij}\nabla ^i\Psi \nabla ^j\Psi $
is identically not vanishing in ${\cal {V}}\setminus \Psi ^{-1}(\Psi _o)$,
since this contradicts that $\Psi |_{{\cal V}}\not \equiv \Psi _o$. If $E$
vanishes in some point of ${\cal {V}}\setminus \Psi ^{-1}(\Psi _o)$, using
its analyticity together with 
$E|_{{\cal {V}}\setminus \Psi ^{-1}(\Psi _o)}\not \equiv 0$, one establishes 
\cite{MullerHRS73,Ayon3} that $E$ vanishes
only in a union of countably many 0-dimensional (point-like), 1-dimensional,
2-dimensional, and 3-dimensional analytic sub-manifolds of 
${\cal {V}}\setminus \Psi ^{-1}(\Psi _o)$. Hence, from (\ref{eq:van}) and the
continuity of $f^{\prime }(\Psi )$ through this lower dimensional
sub-manifolds, it follows that $f^{\prime }(\Psi )$ vanishes identically in 
${\cal {V}}\setminus \Psi ^{-1}(\Psi _o)$, that by (\ref{eq:f}) it requires 
\begin{equation}
\left. \Psi \right| _{{\cal {V}}\setminus \Psi ^{-1}(\Psi _o)}<\Psi _o.
\label{eq:less}
\end{equation}
Let $g\in C^\infty (I\!\!R)$ be a real function defined by 
\begin{equation}
g(t)=\left\{ 
\begin{array}{cr}
-\exp (-1/(\Psi _o-t)^2), & t<\Psi _o, \\ 
0, & t\geq \Psi _o.
\end{array}
\right.  \label{eq:g}
\end{equation}
Since this function satisfies the same properties that $f(t)$, {\em i.e.\/}, 
\begin{equation}
g(\Psi _o)=0,\qquad -1\leq g(t)\leq 0,\qquad \left. g^{\prime }(t)\right|
_{I\!\!R\setminus \Psi _o}\geq 0,  \label{eq:propg}
\end{equation}
it can be concluded, applying the same procedure made to equation 
(\ref{eq:B-Df}) with the function $f$ but this time with $g$, that 
$g(\Psi )$ vanishes identically in 
${\cal {V}}\setminus \Psi ^{-1}(\Psi _o)$. This requires, by (\ref{eq:g}), 
\begin{equation}
\left. \Psi \right| _{{\cal {V}}\setminus \Psi ^{-1}(\Psi _o)}>\Psi _o,
\label{eq:great}
\end{equation}
arriving at a contradiction with (\ref{eq:less}), then the desired result 
$\Psi |_{{\cal {V}}}\equiv \Psi _o\equiv \Psi |_{\langle \!\langle 
\hbox{${\cal J}$\kern -.645em
{\raise.57ex\hbox{$\scriptscriptstyle (\ $}}}\rangle \!\rangle }$ follows,
provided that $T$ is non-positive.

The same result is achieved when $T$ is non-negative; choosing the functions 
\begin{equation}
\hat f(t)=\left\{ 
\begin{array}{cr}
1-\exp (-1/(\Psi _o-t)^2), & t<\Psi _o, \\ 
0, & t\geq \Psi _o,
\end{array}
\right.
\end{equation}
and 
\begin{equation}
\hat g(t)=\left\{ 
\begin{array}{cr}
0, & t\leq \Psi _o, \\ 
\exp (-1/(t-\Psi _o)^2), & t>\Psi _o.
\end{array}
\right.
\end{equation}

Finally we conclude, on the base of the above derived results, that
Brans-Dicke theory of gravitation, with matter satisfying that $T$ has
definite sign, is equivalent to Einstein gravity with the constraint $T=0$,
in presence of static black holes.

\end{document}